\documentclass[pra,twocolumn,showpacs]{revtex4}
\usepackage{graphicx}
\usepackage{dcolumn}
\usepackage{amsmath}
\usepackage{bm}
\renewcommand{\ss}{\scriptscriptstyle}

\newcommand{\sech}{\operatorname{sech}}

\begin{document}


\title{Rabi oscillations of 2D electrons under ultrafast intersubband
excitation}

\author{D. McPeake}
\email{dmcpeake@nmrc.ie}
\author{F. T. Vasko}
\author{E. P. O'Reilly}
\affiliation{NMRC, University College, 
Lee Maltings, Prospect Row, Cork, Ireland }
\date{\today}

\begin{abstract}
We investigate coherent nonlinear dynamics of 2D electrons under ultrafast 
intersubband excitation by mid-IR pulses. We include the effects of relaxation and dephasing, both homogeneous and inhomogeneous, as well as detuning 
 within a non-Markovian equation to obtain temporal population 
redistributions. We show how, using a cross-correlation method, the effects of Rabi oscillations may be detected in this system, and briefly discuss other detection methods. 
\end{abstract}
\pacs{42.50.Md, 73.21.-b, 78.47+p}
\maketitle
The coherent dynamics and manipulation of two-level systems under ultrafast 
excitation dates back over three decades, where the majority of focus has been 
on atomic ensembles \cite{allen}. During recent years, much attention has been given 
to interband coherent effects in low-dimensional heterostructures \cite{rossi}.  
In particular, investigations of single quantum dots under interband ultrafast excitation \cite{gammon}, where dephasing times can be hundreds of picoseconds, has seen the experimental observation of Rabi oscillations, providing the basis for possible readout devices for excitonic quantum gates \cite{nature}.
Previously, interband Rabi oscillations under near-IR excitation had been observed in quantum well (QW) structures, albeit more highly suppressed by shorter dephasing times \cite{rabi1,rabi2}.
Following these interband investigations we investigate the possibility of Rabi oscillations between $n=1$ and $n=2$ conduction subbands in a QW under ultrafast mid-IR excitation.  
Advances in mid-IR femtosecond spectroscopic techniques \cite{isreview} have culminated with
recent experimental results \cite{physica} suggesting that 
`partial Rabi flops' may be possible under relatively low pulse energies. 
 Theoretical investigations of intersubband systems are therefore timely.

The occurrence of subband Rabi oscillations under relatively low pump energies in low-doped QWs is made possible due both to the suppression of the LO-phonon emission rate (assuming the interlevel energy is much greater than the optical phonon energy) and also the reduced efficiency of electron-electron scattering at low densities $(\sim 10^{9}-10^{10}$ cm$^{-2})$. The inclusion of the role played by dephasing mechanisms is crucial for any description of coherent intersubband optics. Whilst a complete treatment requires the full rigour of many-body calculations, we show in this paper that treating the different damping terms phenomonologically within a non-Markovian quantum kinetic scheme gives clear indications of the interplay between the various time-scales involved, and provides insight into the possiblity of coherent manipulation. 
Experimental observations \cite{320prb} show that carrier-carrier dephasing rates are effectively independent of carrier density at low concentration $(\sim 10^{10}$cm$^{-2})$.
We therefore restrict our study to the low-density regime where the assumption of constant dephasing is approximately valid.

After introducing the framework of the two subband model, we consider the balance of population redistributions to derive an integro-differential equation governing population dynamics under the influence of an ultrafast pulse. By considering the ideal limit, where dephasing and relaxation are neglected, we show that our model reduces to analytic expressions which essentially recover the area theorem \cite{mccall}. The effects of the various damping mechanisms are reintroduced in line with recent experimental measurements to investigate the degree of coherent behaviour which can realistically be observed.
We extend our calculations to analyse a detection scheme based on cross-correlation absorption under two-pulse excitation, and briefly discuss a scheme based on THz emission from non-symmetric QWs. 
Such methods have previously been implemented for the near-IR interband pump case \cite{thz,rabi1}. 
We conclude with a brief discussion on the approximations we have made.

We describe the coherent dynamics of the system using a density matrix approach, 
evaluating the balance equation for an interlevel redistribution of
population under a resonant intersubband pump. We consider the response of the
system due to the interaction with an electric field $E_t\equiv [E\exp{(-i\omega t)}+c.c.]w_t$ perpendicular to the QW plane. The dipole
perturbation operator is: $\widehat{\delta h}_t\exp{(-i\omega t)} + H.c.$, 
with $\widehat{\delta h}_t=(ieE/\omega )\hat{\it v}_{\bot}w_t$, where 
$\hat{\it v}_{\bot}$ is the intersubband velocity operator and $w_t$ is  the
form-factor of the pulse. We
write the high-frequency component of the density matrix, $\widehat{\delta{\rho}_{t}}
\exp{(-i\omega t)}+H.c.$, in the form \cite{fedirbook}:   
\begin{equation}
\widehat{\delta\rho}_{t}\simeq \frac{1}{i\hbar} \int_{-\infty}^{0}dt' e^{t'/\tau_2-i\omega t'} e^{(i/\hbar)\hat{h}t'} \left[\widehat{\delta h}_{t+t'} \right] e^{-(i/\hbar)\hat{h}t'} ,
\end{equation}
where $\tau_{2}$, the dephasing time, describes the decay of the induced coherence, and $\hat{h}$ is the period-averaged Hamiltonian of the QW.
The density matrix averaged over the period $2\pi/\omega$, 
$\hat{\rho}_{t}$, obeys the quantum kinetic equation
\begin{equation}
\frac{\partial \hat{\rho}_{t}}{\partial t} + \frac{i}{\hbar}\left[\hat{h} , 
\hat{\rho}_t \right] = \hat{G}_{t} + \hat{I}_{sc} ,
\end{equation}
where 
$\hat{I}_{sc}$ is the collision integral. 
The time-dependent 
generation rate, $\hat{G}_{t}$, is given by:
\begin{eqnarray}
\hat{G}_{t}= \frac{1}{\hbar^{2}} \int_{-\infty}^{0} dt' e^{t'/\tau_{2} -i
\omega t'} \\
\times\left[ e^{i\hat{h}t'/\hbar} \left[\widehat{\delta h}_{t+t'}, 
\hat{\rho}_{t+t'}\right] e^{-i\hat{h}t'/\hbar}, \widehat{\delta h}_{t}^+
\right] + H.c. , \nonumber
\end{eqnarray}
with non-Markovian dependence on $\hat{\rho}_{t}$.
In the resonant approximation, we consider two states $|j\bm{p}\rangle$, $j=1,2$,
which we take to have parabolic dispersion laws, $\varepsilon_{j} + p^2/2m$. 
In this case one can neglect the off-diagonal terms in the density matrix 
under the condition: $\hbar/\varepsilon_{21}\overline{\tau}\ll 1$, where
$\varepsilon_{21}$ is the interlevel energy and $\overline{\tau}$ is the 
characteristic relaxation time. According to particle conservation, we 
have that: $G_{1{\bf p}t}=-G_{2{\bf p}t}\equiv G_{{\bf p}t}$, and the generation rate can be written in the form: 
\begin{equation}
G_{{\bf p}t}=2\left(\frac{e{\rm v}_{\perp}E}{\hbar\omega}\right)^{2}
w_{t} \int_{-\infty}^{0}dt' w_{t+t'}\phi_{t'}\left(f_{2{\bf p}t+t'} - f_{1{\bf p}
t+t'}\right) .
\end{equation}
Here $f_{j\bm{p}t}=\langle j\bm{p}|\hat{\rho}_{t}|j\bm{p}\rangle$ is the population of the $j$th state, $\langle 2|\hat{\it v}_{\bot}|1\rangle =\langle 1|\hat{\it v}_{\bot}
|2\rangle^* =i{\rm v}_{\perp}$ is the interband velocity, $\Delta\omega=\omega 
- \varepsilon_{21}/\hbar$ the detuning frequency, and the kernel is given
by $\phi_t\equiv\cos(\Delta\omega t)\exp{(t/\tau_2)}$. 
Taking into account the inhomogeneous broadening of intersubband transitions
we consider \cite{inhomo} an in-plane nonuniform intersubband energy $\varepsilon_{21}
({\bf x})=\varepsilon_{21}+\delta \varepsilon_{\bf x}$ in the ${\bf x}$-dependent generation rate
$G_{{\bf px}t}$. By averaging over the 2D plane, the detuning term in the generation rate becomes:
$\langle \cos\Delta\omega_{\bf x}t\rangle$. We therefore replace the kernel $\phi_t$ in
Eq. (4) with $\Phi_{t}=\phi_{t}\exp [-(\gamma t /\sqrt{2}
\hbar )^2]$ where $\gamma =\sqrt{\langle\delta\varepsilon_{\bf x}\delta
\varepsilon_{\bf x}\rangle}$ is the energy broadening due to nonuniformity.

Within this framework, we can now define the concentration in the $j$th level as
$n_{jt}= (2/L^2)\sum_{\bm{p}}f_{j\bm{p},t}$; $L^2$ is the normalisation area 
and the factor 2 is due to spin. The balance equations have the form:
\begin{equation}
\frac{d}{dt}\left|\begin{array}{c} n_{1t} \\ n_{2t} \end{array}\right|
=\left| \begin{array}{c}G_t\\-G_t \end{array} \right|+ \left|\begin{array}{c} (\partial n_1/\partial t)_{sc} \\ 
(\partial n_2/\partial t)_{sc} \end{array}\right| ,
\end{equation}
where $G_t=(2/L^2)\sum_{\bm{p}}\langle j\bm{p}|\hat{G}_t|j\bm{p}\rangle$
and $\left(\partial n_{j}/\partial t\right)_{sc} =(2/L^2) \sum_{\bm{p}}
\langle j\bm{p}|\hat{I}_{sc}|j\bm{p}\rangle$ describes interlevel relaxation.
Due to particle conservation, $n_{1t}+n_{2t}=n_{\ss 2D}$ where $n_{\ss 2D}$ is the 
total 2D concentration, so that $\left(\partial n_{1}/\partial t
\right)_{sc}+\left(\partial n_{2}/\partial t\right)_{sc}=0$. Defining the 
population redistribution $\Delta n_{t}= n_{1t}-n_{2t}$, 
and describing the interlevel relaxation by the relaxation time $\tau_1$,
we transform Eq. (5) into 
\begin{equation}
\frac{d\Delta n_{t}}{dt}=2G_{t} + \frac{\Delta n_{t}-n_{\ss 2D}}{\tau_1} ,
\label{integro1}
\end{equation}
where the generation rate $G_{t}$ is expressed through $\Delta n_{t}$
 according to Eq. (4). We solve Eq. (\ref{integro1}) subject to the initial condition $\Delta n_{t\rightarrow -\infty}  = n_{\ss 2D}$.
Upon substitution of Eq. (4) into (6) we obtain
the non-Markovian integro-differential equation
\begin{equation}
\frac{d\Delta n_{t}}{dt} + \nu_{r}w_{t}\int_{-\infty}^{t} \frac{dt'}
{\tau_{p}}w_{t'}\Phi_{t-t'}\Delta n_{t'} + \frac{\Delta n_{t}-n_{\ss 2D}}
{\tau_1} =0 ,
\end{equation}
where we have introduced the phototransition frequency $\nu_{r}=(2e{\rm v}_{\bot}
E/\hbar\omega )^2 \tau_{p}$ with $\tau_{p}$ the pulse duration. Factors which suppress the coherent response
are described by the parameters $(\tau_{1,2})^{-1}$ and $\gamma/\hbar$, as well as detuning of the pulse from resonance $\Delta\omega$. 

Before we resort to numerical solution of Eq. (7), we consider analytic solutions in the limiting case $\tau_{1,2}
\rightarrow \infty$, $\gamma\rightarrow 0$, and $\Delta\omega\rightarrow 0$, ( 
i.e. $\Phi_t \rightarrow 1$). Eq. (7) can then be transformed into the 
second order differential equation
\begin{equation}
\frac{d^{2}\Delta n_{t}}{dt^{2}}- \frac{1}{w_{t}}\frac{dw_{t}}{dt}
 \frac{d \Delta n_{t}}{dt}
+\frac{\nu_r}{\tau_p}w_{t}^{2} 
\Delta n_{t} = 0 ,
\end{equation}
to which we apply an additional boundary condition: $\left[w_{t}^{-1}d\Delta n_{t}/dt
\right]_{t\rightarrow -\infty}=0$. 
The population dynamics in this regime can be described by the expression
\begin{equation}
\Delta n_{t} = \cos\left(\sqrt{\nu_{r}\tau_{p}}\int\limits_{-\infty}^{t/
\tau_p}dz w_{z} \right) ,
\end{equation}
which we illustrate in Fig. 1.
As expected, population redistributions are solely determined by the area of the incident pulse: $A_p = \sqrt{\nu_{r}\tau_{p}} \int_{-\infty}^{\infty} dz w_{z}$. In Fig. 1a we plot the solution as a function of time for both secant and Gaussian pulses each with duration $\tau_{p}$ (FWHM). As illustrated in panels (i) and (ii), which correspond to secant pulses with $A_{p}=\pi$ and $2.5\pi$ respectively, population can be distributed among the two levels according to the area theorem. For fixed values of $\nu_{r}$ and $\tau_{p}$, the area under the Gaussian pulse is slightly less than that under the secant pulse. 
The final redistribution of population, shown in Fig. 1b as a function of the secant pulse area, $\pi\sqrt{\nu_{r}\tau_{p}}/\alpha_{s}$, where $\alpha_{s}=2\sech^{-1}[0.5]$, confirms this point.

\begin{figure}
\begin{center}
\includegraphics[width=7cm,bb= 8 33 706 411,clip=]{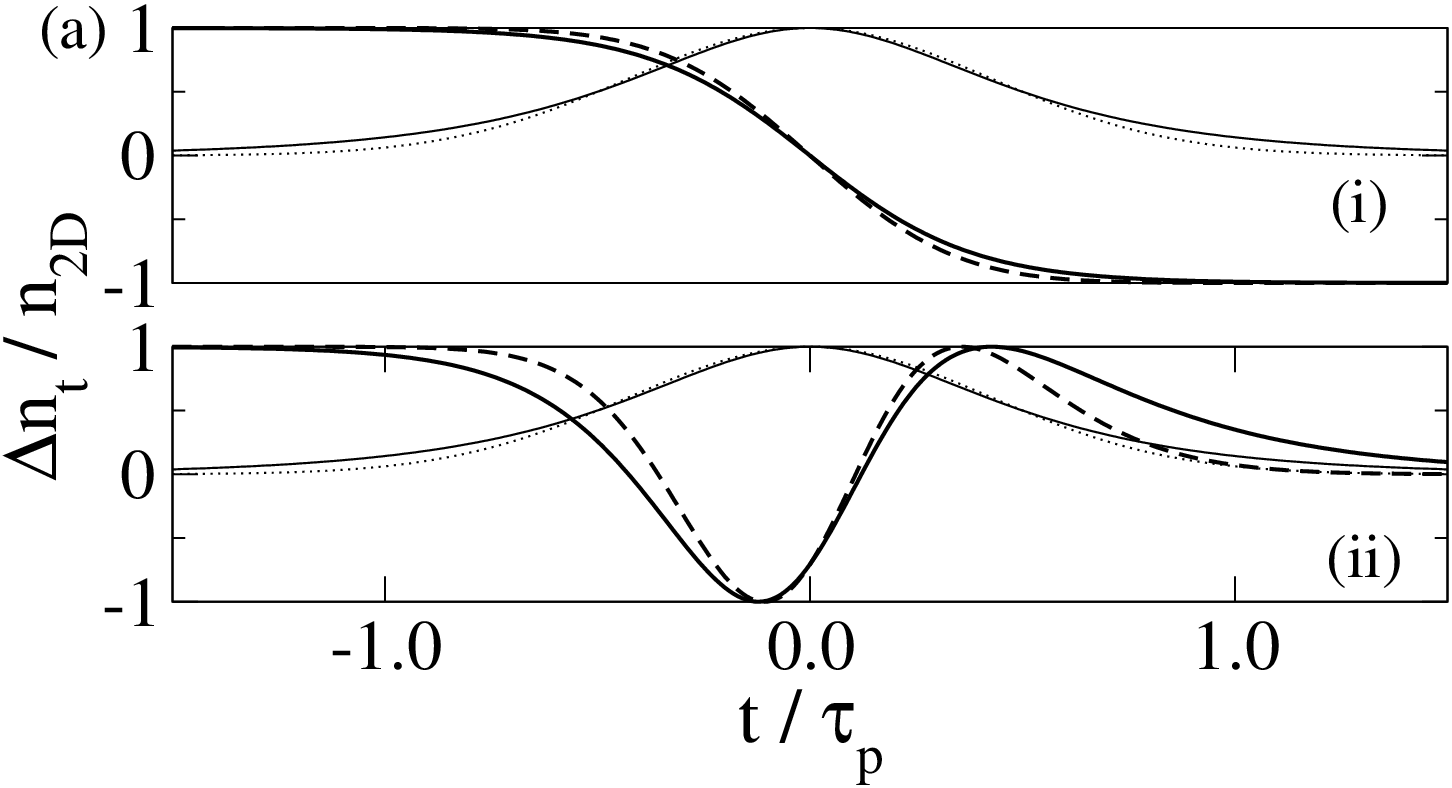}
\includegraphics[width=7cm,bb= 12 8 724 303,clip=]{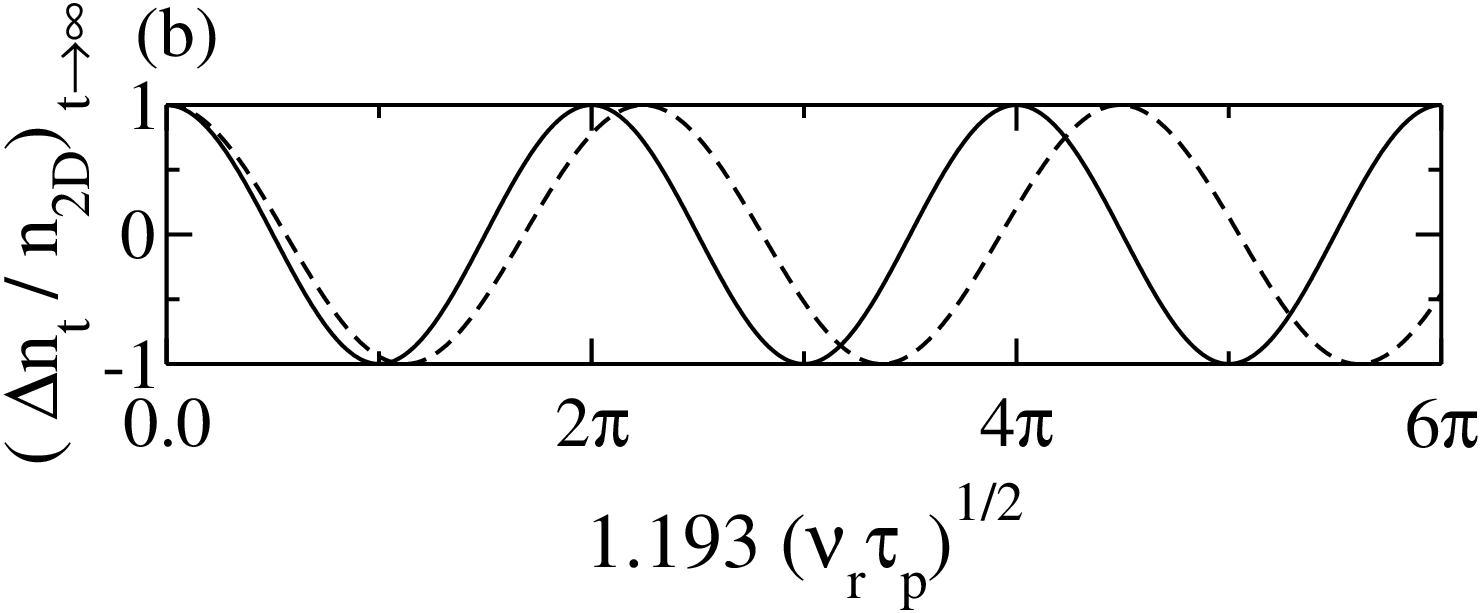}
\end{center}
\addvspace{-0.5 cm}
\caption{(a) Temporal population redistribution, $\Delta n_t$, for the ideal case ($\tau_{1}\rightarrow \infty, \Phi_{t}\rightarrow 1$) due to secant (thick solid curve) and Gaussian (dashed curve) pulses of duration $\tau_{p}$ (FWHM). The pulse profiles are shown as the thin solid (secant) and dotted curves  (Gaussian). Panels (i) and (ii) correspond to pulse areas (for secant pulse) $A_{p}= \pi$ and $2.5\pi$ respectively. (b) The area theorem is recovered as final distribution is plotted against secant pulse area ($1.193\sqrt{\nu_{r}\tau_{p}}$) for secant (solid curve) and Gaussian (dashed curve) profiles. 
}
\end{figure}

The condition for population inversion, $A_{p}=\pi$
can be obtained by a secant pulse, with duration $\tau_p=$100 fs, having a free space intensity per unit area of 5.22 MW/cm$^{2}$. 
In order to achieve maximum coupling of the transverse field to the intersubband dipole, we consider a 45$^{\circ}$ prism integrated onto the sample. With this geometry, approximately 33$\%$ of the incident power can be coupled to the transition.
 We assume parameters for a typical GaAs QW, of width 85 \AA, 
corresponding to an interlevel energy of $\hbar\omega\simeq$100 meV. 
Both sets of parameters are typical of recent experiments \cite{isreview}.

To study the effects of the various damping processes, we solve Eq. (7) numerically using a form of Picard iteration. We take as example a secant
 1.6$\pi$-pulse, and consider the effects of the parameters, $\tau_1$, $\tau_2$, $\Delta\omega^{-1}$, and $\sqrt{2}\hbar /\gamma$, setting each in turn to $\tau_p$ with the other three set to the limiting values as outlined above. The results are shown in Fig. 2. LO-phonon emission ($\tau_{1}$) causes slow relaxation of the population back to the ground state after the pulse has passed. In contrast, the effect of a large detuning as well as broadenings, both inhomogeneous and homogeneous ($\sqrt{2}\hbar/\gamma$, $\tau_2$), is to cause suppression of the Rabi flop during the pulse. Clearly the potential for coherent manipulation depends on the interplay between these processes.

To include realistic estimates of these parameters, we take experimentally determined values of coherence times for a typical sample \cite{physica,320prb}. Four-wave-mixing measurements provide a value for the homogeneous broadening corresponding to an electron-electron scattering time of the order 320 fs.
LO-phonon emission times are estimated to be between 1-4 ps. As a conservative estimate we take $\tau_{1}=1$ ps and calculate the system response to a 100 fs (FWHM) pulse tuned to the $1\rightarrow 2$ transition at $E$=100 meV, and allow for a detuning of up to 12 meV. In Figs. 3a,b we show the population redistributions due to these parameters.
\begin{figure}
\begin{center}
\includegraphics[width=7cm,bb= 8 3 726 533,clip=]{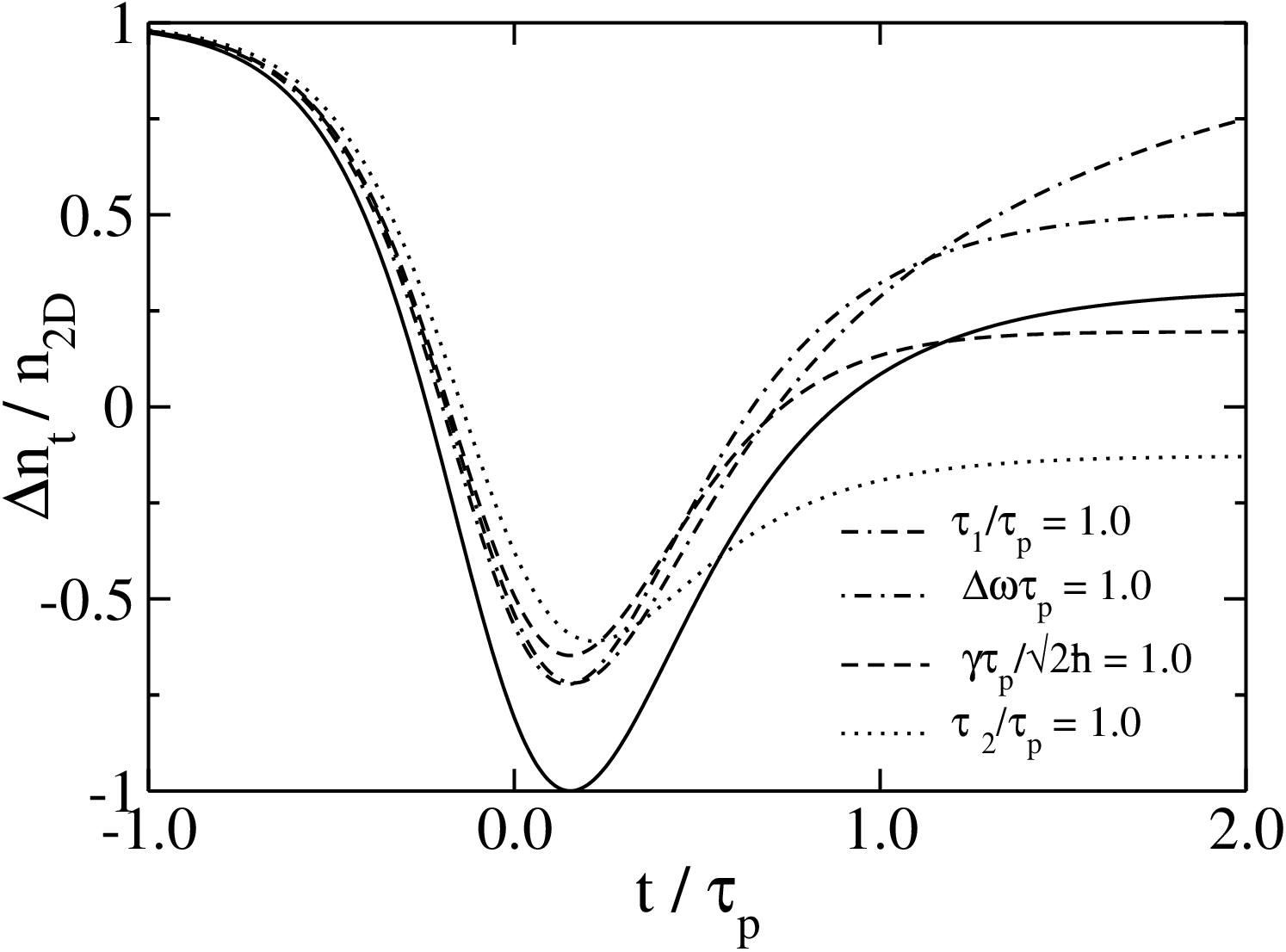}
\end{center}
\addvspace{-0.5 cm}
\caption{ The suppression of the ideal Rabi flop (solid curve) under 1.6$\pi$-pulse excitation (duration $\tau_{p}$ (FWHM)) due to: detuning (dot-dashed curve), inhomogeneous broadening (dashed curve), homogeneous broadening (dotted curve) and LO-phonon emission (dot-double-dashed curve). Each parameter is varied in turn with the rest set to ideal values.}
\label{exact_sol}
\end{figure}
\begin{figure}
\begin{center}
\includegraphics[width=7cm,bb= 2 153 725 594,clip=]{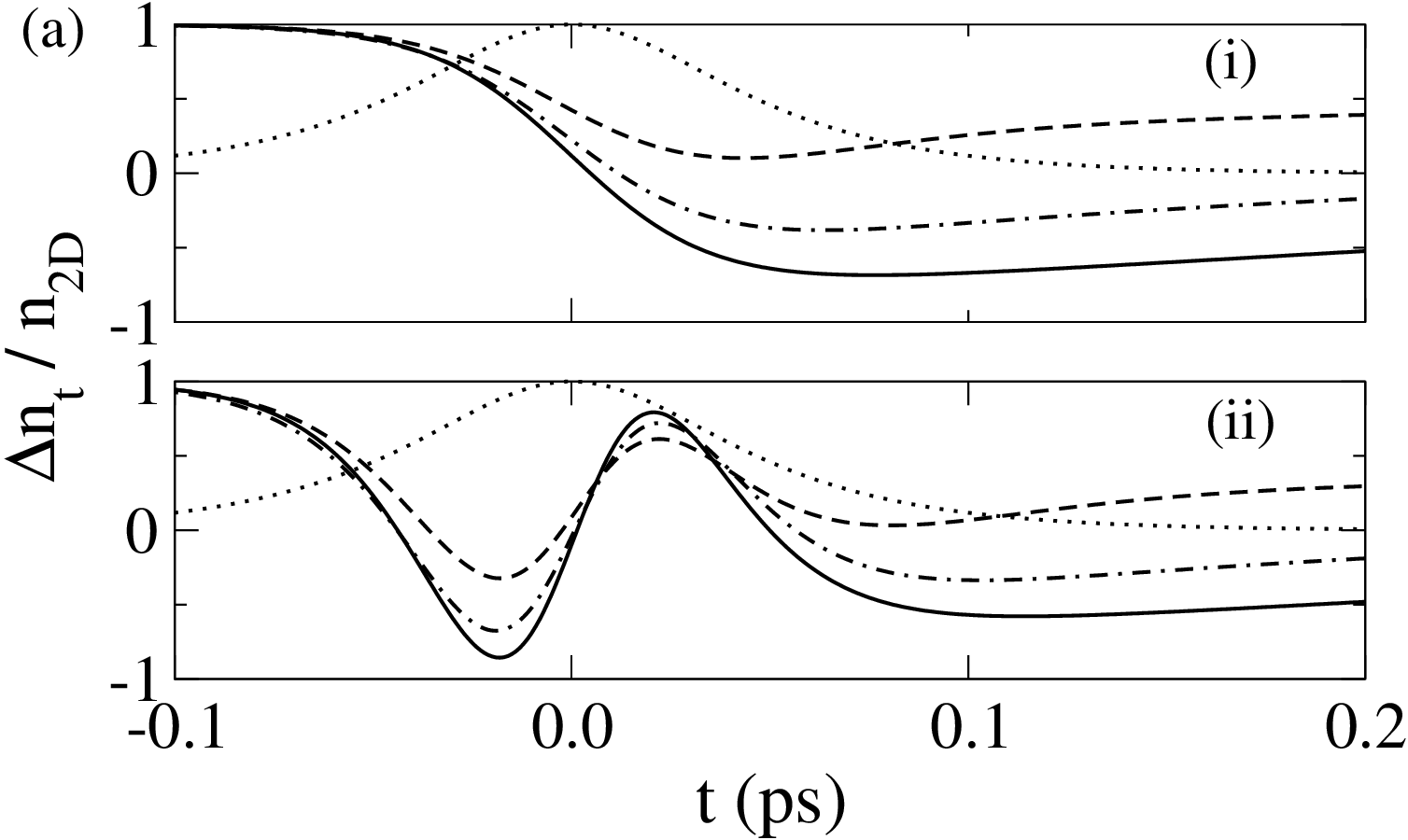}
\includegraphics[width=7cm,bb= 31 8 780 299,clip=]{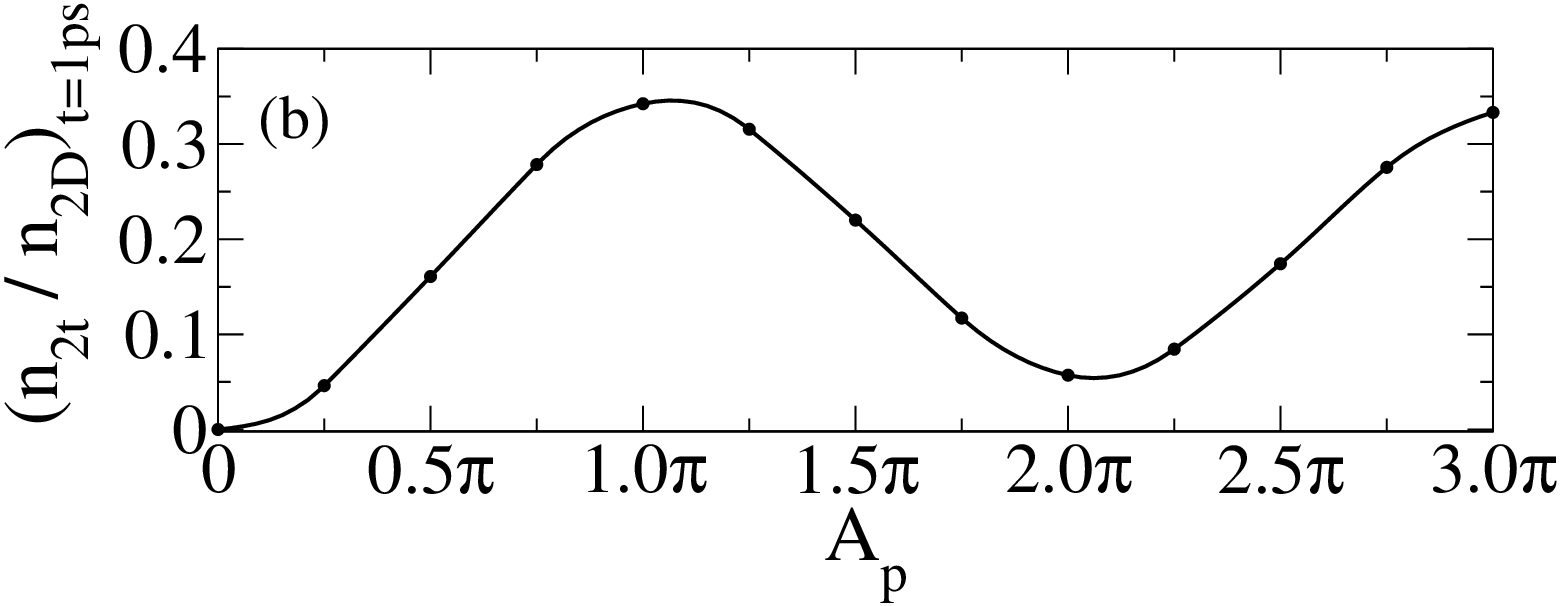}
\end{center}
\addvspace{-0.5 cm}
\caption{(a) Temporal population redistributions
between subbands levels 1 and 2 due to excitation by a 100 fs (FWHM) secant pulse of area (i) $\pi$, (ii) $3\pi$. Decay parameters are taken to be $\tau_{2}=320$ fs, $\tau_{1}=1$ ps, corresponding to a two-level system homogeneously broadened by 4 meV. Also shown is the effect of detunings of 6 meV (dot-dashed curves) and 12 meV (dashed curves). The fraction of population excited at 1 ps after the maximum pulse intensity is shown as a function of $A_p$ in (b).} 
\label{3}
\end{figure}
\begin{figure}
\begin{center}
\includegraphics[width=7cm,bb= 42 13 727 522,clip=]{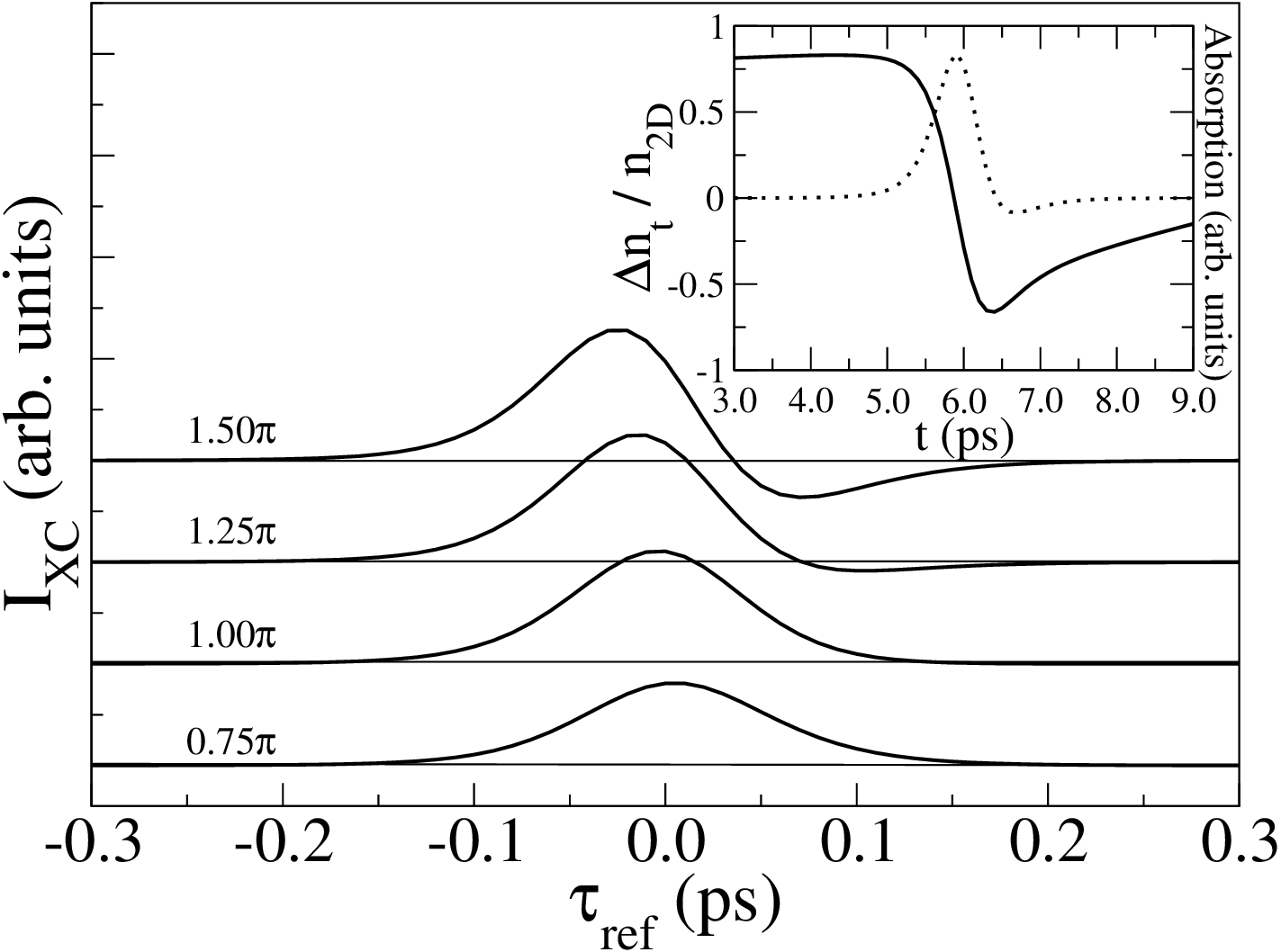}
\end{center}
\addvspace{-0.5 cm}
\caption{Differential cross-correlations, shifted vertically for clarity, for different pulse areas (numbers to the left of peaks) as a function
of reference time delay, $\tau_{\ss ref}$.
Inset: Temporal absorption (dotted
curve) and density
variation (solid curve) for the pulse with area $A_{p}=1.25\pi$.}
\label{4}
\end{figure}
Although complete inversions are suppressed, we estimate that up to 70-80\% of the initial distribution can be coherently excited to the upper level with a resonant $\pi$-pulse: Fig. 3a (i). Effects of pulse area can still be seen in the population 1 ps after the maximum of the pulse intensity. However the question remains as to whether these density flops can be detected experimentally. We now consider two possible scenarios. 

Following the interband example \cite{rabi1}, we calculate cross-correlation functions comparing changes in pulse modification due to transmission through unexcited and weakly excited samples. 
With the
induced current density written in the standard form: $J_t=(2e/L^2)
\sum_{jj'{\bf p}}{\rm v}_{jj'}\langle j'{\bf p}|\widehat{\delta\rho}_{t}|j
{\bf p}\rangle$, we can use Eq. (1) to express the induced current through
$\Delta n_t$. The absorbed power, ${\cal P}_{t}=\overline{J_tE_t}$ where the
overline means the average over the period, is written in the form:
\begin{equation}
{\cal P}_t=\frac{\left(2e{\rm v}_{\bot}E\right)^{2}}{\hbar\omega}w_t\int_{-\infty}^{0}dt'w_{t'}
e^{t'/\tau_2}\cos{\Delta\omega t'}\Delta n_{t+t'} .
\end{equation}
The transmitted intensity of a single pulse is thus given by: $I_{s}=I_{t}-{\cal P}_{t}$, where $I_{t}\propto |E_{t}|^2$. $I_{d}$, the transmitted intensity of an identical pulse after weak excitation of the sample by a prepulse, is similarily extracted from Eqs. (7)(10) using the two-pulse form-factor, $w_{t}\rightarrow w_{t}^{pre} + \beta w_{t}$ where $\beta=E/E_{pre}$ is the ratio of the maximum of the electric field of the pulse and prepulse. The cross-correlation is thus:
\begin{equation}
I_{\ss XC}(\tau_{\ss ref}) \propto \int dt\left[I_{d}(t) - I_{s}(t)\right] I_{r}(t-\tau_{\ss ref}) ,
\end{equation}
where $I_{r}(t-\tau_{\ss ref})$ is the intensity of a reference pulse (we take $I_{r}(t-\tau_{\ss ref}) \propto |E_{t}|^2$). The interval between pulse and prepulse is kept fixed at 600 fs allowing for no coherent interaction between pulses. The prepulse area is fixed at $A_{p}^{(pre)}=0.25\pi$. The intensity (area) of the pulse and hence reference pulse is varied and the cross-correlation (11) calculated. The results are shown in Fig. 4.
The appearance of zero crossings in the cross-correlation, which move to shorter time-delays with increasing pulse intensity, is exactly the behaviour expected due to Rabi flops. These crossings which occur due to the dependence of the cross-correlation signal on the temporal derivative of the excitation density during the pulse, show that the system has been driven through a density maximum - a partial Rabi flop.             

In addition to this cross-correlation scheme, we envisage detection based on the charge oscillations of a Rabi flopping population in a non-symmetric QW with dipole moments $d_1$ and $d_2$. The temporal dependence of the induced dipole moment, $n_{1t}d_{1}+n_{2t}d_{2}\propto (1-{d_2}/{d_1})\Delta n_t$, connects the temporal population dynamics, $\Delta n_t$, with the induced oscillations of the electric field. The response can be measured by the method employed in \cite{physica}, as well as by the emission of THz radiation. We note that the experimental method \cite{thz} used for the detection of THz radiation under interband excitation was carried out with carrier densities comparable to those we have considered.       

Our calculations are based on a few assumptions. Since the Coloumb renormalisation of the intersubband absorption (due to depolarisation and exchange effects) is weak in low-doped QWs \cite{320prb,inhomo}, we have adopted a single-particle approach. All homogeneous relaxation and dephasing mechanisms including electron-electron scattering and LO-phonon emission have been considered phenomenologically with inhomogeneous broadening included in the long-scale limit. We expect that microscopic considerations based on many-body calculations of the various damping processes, taking into account in-plane inhomogeneities, may slightly modify the nonlinear response. However, we do not expect large departures from this simple model for the low-density regime we have considered. 
Due to the prism geometry we have assummed, the finite length of the pulse, $L$, must be taken into account. For typical experimental parameters we estimate that $L\approx 15$ $\mu$m, thus setting a limit on the sample area, and hence amplitude of response we can describe. However, we note that the temporal dynamics we have described are not changed as $L$ typically exceeds the characteristic kinetic lengths, $\overline{v}\tau_{1,2}$ and $\overline{v}/\nu_{r}$, where $\overline{v}$ is the characteristic in-plane velocity.

In conclusion, we have developed a simple description of Rabi oscillations between subband levels in low-doped QWs. We have considered the population dynamics of a two-level system under ultrafast mid-IR radiation in a non-Markovian quantum kinetic model.   
Our calculations, based on realistic pulse and material parameters, show
that coherent manipulation of the order of 10$^{6}$ electrons (depending on area of excitation) should be possible under relatively low-power excitations.  

This work was funded by Science Foundation Ireland. 

\end{document}